\documentclass[12pt]{article}
\usepackage[dvips]{color}
\usepackage{pennames,epsfig}

\def\babar{{\sl B\hspace{-0.4em} {\scriptsize\sl A}\hspace{-0.4em}
B\hspace{-0.4em} {\scriptsize\sl A\hspace{-0.1em}R}\,\,}}

\def\ifmath#1{\relax\ifmmode#1\else$#1$\fi}

\newcommand{\comment}[1]{}

\def\vs{{\it vs.}}

\def\CP{\ifmath{C\!P}}

\def\BF{$B$ Factory}

\def\Bmeson{$B$ meson}
\def\Bmesons{$B$ mesons}

\def\babar{\mbox{\sl B\hspace{-0.4em} {\scriptsize\sl A}\hspace{-0.4em} B\hspace{-0.4em} {\scriptsize\sl A\hspace{-0.1em}R}}}

\def\dirc       {{\sc DIRC}}

\def\PEPII	{\mbox{PEP-II}}

\def\bz    {\ifmath{B^0}}
\def\bzb   {\ifmath{\overline{B}{}^0}}

\def\BzBzb {\ifmath{B^0 \overline{B}{}^0}}

\def\Y#1S  {\ifmath{\Upsilon (#1S)}}

\def\ctheta	{\mbox{$\cos \theta$}}

\def\kev  {\ifmath{\mbox{\,ke\kern -0.08em V}}}
\def\mev  {\ifmath{\mbox{\,Me\kern -0.08em V}}}
\def\gev  {\ifmath{\mbox{\,Ge\kern -0.08em V}}}
\def\gevc {\ifmath{\mbox{\,Ge\kern -0.08em V$\!/c$}}}
\def\mevc {\ifmath{\mbox{\,Me\kern -0.08em V$\!/c$}}}
\def\gevcc{\ifmath{\mbox{\,Ge\kern -0.08em V$\!/c^2$}}}
\def\mevcc{\ifmath{\mbox{\,Me\kern -0.08em V$\!/c^2$}}}

\def\m    {\ifmath{\mbox{\,m}}}

\def\cm   {\ifmath{\mbox{\,cm}}}

\def\mm   {\ifmath{\mbox{\,mm}}}

\def\mum  {\ifmath{\,\mu\mbox{m}}}

\def\nm   {\ifmath{\,\mbox{nm}}}

\def\ns   {\ifmath{\,\mbox{ns}}}

\def\mrad {\ifmath{\mbox{\,mrad}}}
\def\rad {\ifmath{\mbox{\,rad}}}
\def\deg  {\ifmath{^\circ}}

\makeatletter
\def\@versim#1#2{\vcenter{\offinterlineskip
        \ialign{$\m@th#1\hfil##\hfil$\crcr#2\crcr\sim\crcr } }}
\def\gsim{\mathrel{\mathpalette\@versim>}}
\def\lsim{\mathrel{\mathpalette\@versim<}}
\makeatother

\def\protoI	{\mbox{Prototype I}}  
\def\protoII	{\mbox{Prototype II}}   
\def\SOB	{\mbox{Standoff Box}} 
\def\PMT     {{\sc PMT}}

\def\rms     {root mean squared}
\def\gau {Gaussian}
\def\photoelectron {photoelectron}

\def\thc{\ifmath{\theta_C}}
\def\phc{\ifmath{\phi_C}}
\def\dip{\ifmath{\theta_D}}

\def\ev  {\ifmath{\mbox{\,e\kern -0.08em V}}}

\def\babar{{\sc BaBar}}
\def\dirc{{\sc Dirc}}
\def\PEPII{{\sc Pep-II}}
\def\BF{\mbox{B Factory}}
\def\BzBzb {\ifmath{{\rm B}^0 \overline{{\rm B}}{}^0}}
\def\BBb {\ifmath{{\rm B}\overline{\rm B}}}
\def\Bmeson{B meson}
\def\Bmesons{B mesons}
\def\bz    {\ifmath{{\rm B}^0}}
\def\bzb   {\ifmath{\overline{{\rm B}}{}^0}}
\newcommand{\cer} {Cherenkov}

\input epsf

\begin{document}

\thispagestyle{empty}
\renewcommand{\thefootnote}{\fnsymbol{footnote}}

\begin{flushright}
{\small
SLAC--PUB--7715\\
December 1997\\}
\end{flushright}

\vspace{.8cm}

\begin{center}
{\bf\large   
\dirc , a New Type of Particle Identification System for 
\babar \footnote{Work supported by
Department of Energy contract  DE--AC03--76SF00515.}}

\vspace{1cm}

\vspace{1.cm}

{\bf Jochen Schwiening}

Stanford Linear Accelerator Center

Stanford University, Stanford, CA 94309

\vspace{3.mm}
{\it Representing}
\vspace{.2cm}

{\bf The \babar -\dirc\ Collaboration}

\end{center}

\vfill

\begin{center}
{\bf\large   
Abstract }
\end{center}

\begin{quote}
The \dirc , a new type of \cer\ imaging device, 
has been selected as the primary particle identification system for 
the \babar\ detector at the asymmetric B-factory, {\sc Pep-II}.
It is based on total internal reflection and uses long, rectangular bars
made from synthetic fused silica as \cer\ radiators and light guides.
In this paper, the principles of the {\dirc} ring imaging \cer\ technique are
explained and results from the prototype program are presented.
The studies of the optical properties and radiation hardness of the 
quartz radiators are described, followed by a discussion of the 
detector design.
\end{quote}

\vfill

\begin{center} 
{\it Invited talk presented at the} \\
{\it 5th International Workshop on B--Physics at Hadron Machines (Beauty'97)}\\
{\it Santa Monica, California, USA}\\
{\it October 13--17, 1997}\\
\end{center}

\newpage



%
\pagestyle{plain}

\section{Introduction}
{\PEPII} is an asymmetric \Pep \Pem\ collider, with beam energies of 9 GeV
electrons upon 3.1 GeV positrons~\cite{pep2}.
At the design luminosity of $10^{34}$ cm$^{-2}$s$^{-1}$, the production of
\PgUc\ with a boost of $\beta\gamma = 0.56$ will result in about 10 {\BBb}
pairs per second.
{\babar} is the detector dedicated to studying the
collisions at {\PEPII}, with a primary physics goal of observing
{\CP} violation in the {\BzBzb} system~\cite{tdr}.
The experiment is expected to begin taking data in the spring of 1999.

In the study of {\CP} violation, precise particle identification (PID) of
charged pions and kaons over the full kinematic range is of particular
importance, both to reconstruct one of the two
{\Bmesons} in an exclusive decay mode, and to ``tag'' the beauty content of
the recoiling {\Bmeson} (identify it as either a {\bz} or a {\bzb}
when it decayed).
The \babar\ drift chamber can perform \Pgppm /\PKpm\ separation with at least
3 $\sigma$ significance up to 700 MeV/c by measuring the specific energy loss.
Since the B-decays are boosted at {\PEPII} in the forward (electron)
direction, the maximum momentum particle possible
in a two-body B-decay is about 1.5 {\gevc} in the backward direction and
about 4 {\gevc} in the forward direction.
A dedicated PID system should cover the range from 0.7 up to 4 {\gevc} and
take the asymmetry of the boosted events into account.
Since the PID system is surrounded by a CsI crystal calorimeter, it should
also be thin in both radial dimension (to minimize the amount of CsI
material needed) and radiation length (to avoid the deterioration of the
excellent energy resolution of the calorimeter).
Finally, to operate successfully in the high-luminosity environment of {\PEPII},
the detector should be fast and tolerant of background.

\section{The {\dirc} Concept}
\label{sec:principle}

The \dirc\ is a new type of \cer\ ring imaging detector
which utilizes totally internally reflecting \cer\ photons
in the visible and near UV range~\cite{dirc_concept}.
The acronym \dirc\ stands for {\bf D}etection of {\bf I}nternally
{\bf R}eflected {\bf{\v{C}}}erenkov light.

\begin{figure}
  \hspace*{-2mm}
    \mbox{\epsfig{figure=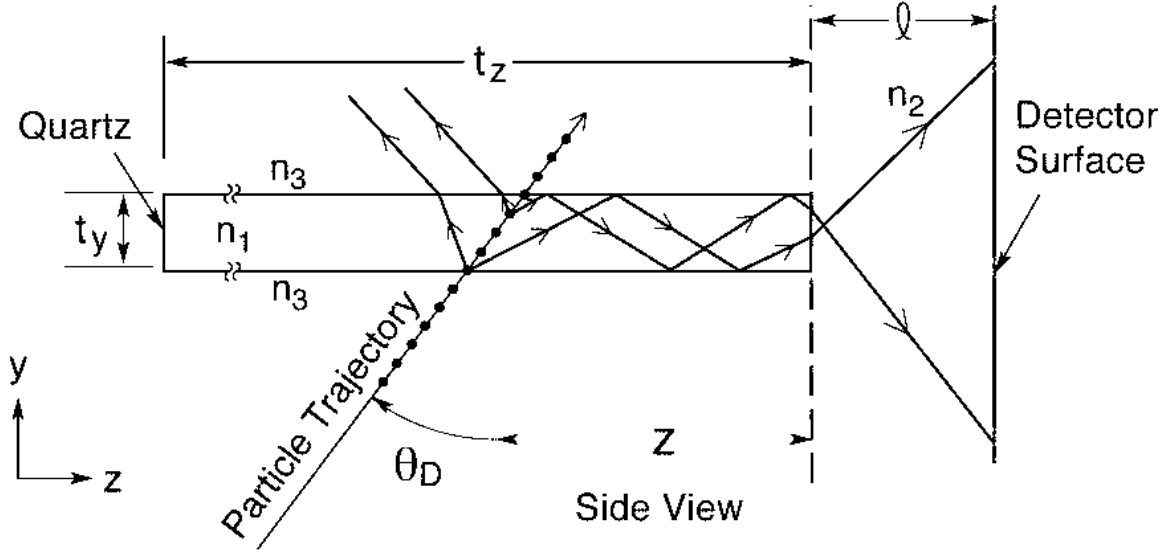,width=\textwidth}}
  \caption
    {\it \label{fig:concept}
	Imaging principle of the \dirc .
    }
\end{figure}

The geometry of the \dirc\ is shown schematically in Figure~\ref{fig:concept}.
It uses long, thin, flat quartz radiator bars
(effective mean refractive index $n_{1}$ = 1.474)
with a rectangular cross section.
The quartz bar is surrounded by a material
with a small refractive index $n_{3} \sim 1$ (nitrogen in this case).
As it traverses the quartz bar, a particle of velocity
\mbox{$\beta  = \frac{v}{c} \ge 1/n_{1}$} will
radiate \cer\ photons in a cone of half opening angle $\theta_c$ around
the particle trajectory.
Since the refractive index of the radiator bar $n_{1}$ is large,
some of the \cer\ photons will be totally internally reflected,
regardless of the incidence angle of the tracks, and propagate
along the length of the bar.
To avoid having to instrument both bar ends with photon detectors, a mirror
is placed at one end, perpendicular to the bar axis.
This mirror returns most of the incident photons to the other (instrumented)
bar end.
Since the bar has a rectangular cross section and is made to optical precision,
the direction of the photons remains unchanged and the \cer\ angle conserved
during the transport, except for left-right/up-down ambiguities due to the
reflection at the radiator bar surfaces.
The photons are then proximity focused by expanding through
a standoff region filled with purified water (index $n_{2} \sim 1.34$)
onto an array of densely packed photomultiplier tubes
placed at a distance of about 1.2 m from the bar end, where the \cer\
angle is measured from the radius of the \cer\ ring, determining the 
particle velocity.
Combined with the mo\-men\-tum information from the drift chamber, the
mass of the particle is identified.

\section{The {\dirc} Prototype Program}
\label{sec:dirc_proto}

The topics described here were selected to highlight the
performance of the \dirc\ prototypes.
Details of the setup, data analysis, and results can be found in
Ref.~\cite{hide_ieee} for {\protoI} and in Ref.~\cite{nim_paper} for
{\protoII}.

\subsection{\protoI}

{\protoI} was a conceptual prototype that operated from 1993--1994
in a hardened cosmic muons setup at SLAC~\cite{hide_ieee}.
It provided a proof-of-principle and the basis for the detector simulation
and performance estimates for the \dirc\ in \babar .
Its main goal besides proving the feasibility of the \dirc\ concept
was to measure the photon yield and single \cer\ photon angle resolution.

The setup 
contained 120~cm and 240~cm long quartz bars (manufactured by
Zygo Corp.~\cite{zygo} from Vitreosil~F/055 fused quartz material~\cite{qpc}),
4.73~cm wide and 1.70~cm thick.
The 240~cm long bar was made from two 120~cm long bars
glued together with an epoxy (epo-tek 301-2~\cite{epotek})
which has good transmission at wavelengths to which the PMTs are sensitive.
The quartz bars were placed in a light-tight box and supported by nylon screws.
The box was mounted on rotating rails
to allow angle and position variation.
The cosmic muon trigger was provided by scintillation counters,
and a 1 {\gevc} threshold was provided by an iron stack underneath the bar.
An array of straw chambers was used to measure the track direction
for the single \cer\ photon angle resolution measurement.

\begin{figure}
  \begin{center}
    \mbox{\epsfig{figure=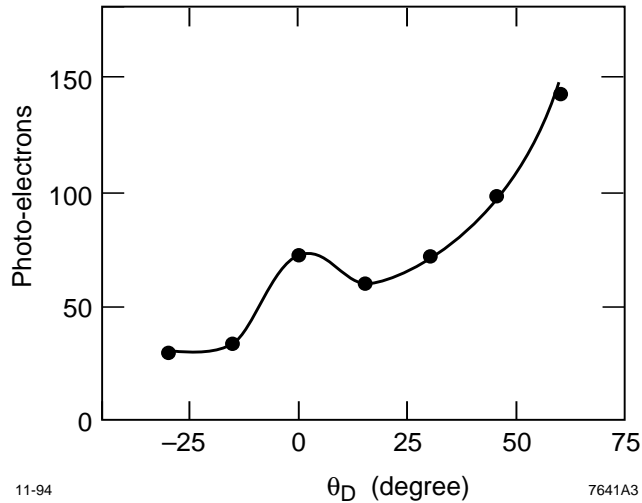,width=9.cm}}
  \caption
    {\it \label{fig:p1angle}
	The observed photoelectron yield from a 1.7~cm thick bar
	at a track position $z$ = 60 cm
	as a function of dip angle $\theta_{D}$.
	The solid line is a Monte Carlo simulation.
	The statistical errors of the measurements
	are smaller than the circles.
	There is a scale error of 3~\% due to calibration uncertainty.
    }
  \end{center}
\end{figure}

To measure the photoelectron yield,
a single 2$^{\prime\prime}$ diameter PMT
(Burle-8850~\cite{pmtp1})
was glued directly to the end of the 120~cm long bar.
Figure~\ref{fig:p1angle}
shows the observed average number of photoelectrons
as a function of track dip angle.
The number of photoelectrons can be written using
the ``\cer\ quality factor''  $N_{0}$ as
\begin{equation}
        N_{pe} = \epsilon_{coll} 
		 \frac{d}{cos \theta_{D}}
		 N_{0} sin^{2}\theta_{C},
\end{equation}
where $\epsilon_{coll}$ is the photon collection efficiency,
$\theta_{C}$ is the \cer\ angle,
$d$ is the thickness of the radiator bar,
and $\theta_{D}$ is the dip angle of the track.
The collection efficiency $\epsilon_{coll}$ is a function of
dip angle, track position, and photon energy.
The $N_{0}$ is given by
\begin{equation}
	N_{0} = \frac{\alpha}{\hbar c}\int \epsilon_{PMT} dE,
\end{equation}
where $\epsilon_{PMT}$ is the quantum efficiency of the PMT 
and $E$ is the photon energy. 
From the quantum efficiency curve claimed by the manufacturer,
we find $N_{0} \sim$~ 150~cm$^{-1}$.
The solid line is a Monte Carlo simulation of the test setup.
It simulates the propagation of the photons through the bar,
taking into account
the wavelength dependencies of both the photon absorption in the quartz bar
and the quantum efficiency of the PMT.
The Monte Carlo simulation reproduces
both the dip angle dependence and absolute yield well.
The number of photoelectrons increases in the
forward direction, mainly as a result of the larger amount of quartz
traversed by the particles.
The bump at $\theta_{D}~=~0^{\circ}$ is due to the fact
that all of the photons are internally reflected at this angle.

To measure the angular resolution,
a closely packed array of 47 1$\frac{1}{8}$$^{\prime\prime}$ diameter PMTs 
(EMI-9124A~\cite{emi_tubes}) was used.
The standoff region ({\SOB}) material was air to allow easy movement of the
array.
The single \cer\ photon angle resolution was meaured with the PMT array
at various standoff distances~$\ell$, dip angles $\theta_{D}$,
and track positions $z$.
The results of the measurements are summarized in Table~\ref{tab:resolutions}.
The quoted values for the resolution are from a fit of a Gaussian plus
polynomial to the \cer\ photon angle distribution.
The measured single photon resolution
is found to be consistent with the estimate based on the Monte Carlo
simulations and no significant dependence on the track position is seen.

\begin{table}
  \begin{center}
    \mbox{
      \begin{tabular}{|c|c|c||c|c|} \hline
	& & & \multicolumn{2}{c|}{resolution [mr]} \\ \cline{4-5}
	$\theta_{D}$ & $\ell$ [cm] & $z$ [cm]
	& measurement & estimate
	\\ \hline \hline
	30$^{\circ}$ & 60 &  60 & $10.1\pm1.1$ & 10.1 \\ \hline
	40$^{\circ}$ & 60 &  60 & $11.7\pm1.2$ & 12.0\\ \hline
	50$^{\circ}$ & 60 &  60 & $14.0\pm0.9$ & 12.8 \\ \hline
	60$^{\circ}$ & 60 &  60 & $12.6\pm0.8$ & 12.5 \\ \hline \hline
	60$^{\circ}$ & 90 &  30 & $ 8.6\pm1.4$ & 10.2 \\ \hline
	30$^{\circ}$ & 90 &  30 & $ 8.5\pm0.5$ &  7.8 \\ \hline
	30$^{\circ}$ & 90 & 110 & $ 7.5\pm1.2$ &  7.8 \\ \hline \hline
	30$^{\circ}$ & 90 & 205 & $ 8.5\pm0.6$ &  7.8 \\ \hline
      \end{tabular}
    }
  \caption
    {\it \label{tab:resolutions}
	Summary of the single \cer\ photon resolution measurements.
	The measurement at $z$ = 205~cm is for the 240~cm long bar.
	All others are for the 120~cm long bar.
    }
  \end{center}
\end{table}

In other tests, the \cer\ photon attenuation rate along the bar
was found to be $\sim$ 10\%/m for the track dip angle
\mbox{$\theta_{D} = 30 \deg$}
and no significant light loss or resolution loss
at the glue joint in the 240~cm long bar was seen.

In conclusion, all measurements were consistent with each other
and with the Monte Carlo simulation and estimates.
They demonstrated the particle identification capability of the \dirc ,
and showed that the main features of the device were well-understood
and that the performance could be safely extrapolated to a full scale device.

\subsection{\protoII}

After the
completion of {\protoI}, many issues remained to be
resolved before a {\dirc} detector for {\babar} could be designed.
The full scale {\protoII}~\cite{nim_paper} was constructed to address
these concerns.
The main goals were:
\begin{itemize}
	\item to refine the early performance estimates;
	\item to explore the engineering issues associated with
	constructing a large {\dirc} detector;
	\item to gain experience in the long-term operation of a large
	{\dirc} detector;
	\item to provide a test-bench for new ideas as the design of the
	{\babar} {\dirc} proceeded.
\end{itemize}
The construction of the prototype was completed in April 1995, and it
was tested briefly in a hardened cosmic muon beam at LBNL.
Shortly thereafter, the detector was shipped to CERN where it was installed in
the T9 zone of the CERN PS East Hall.
It was tested for a period of over 12 months in the T9 beam.

\begin{figure}
  \vspace*{-5mm}
  \begin{center}
    \mbox{\epsfig{figure=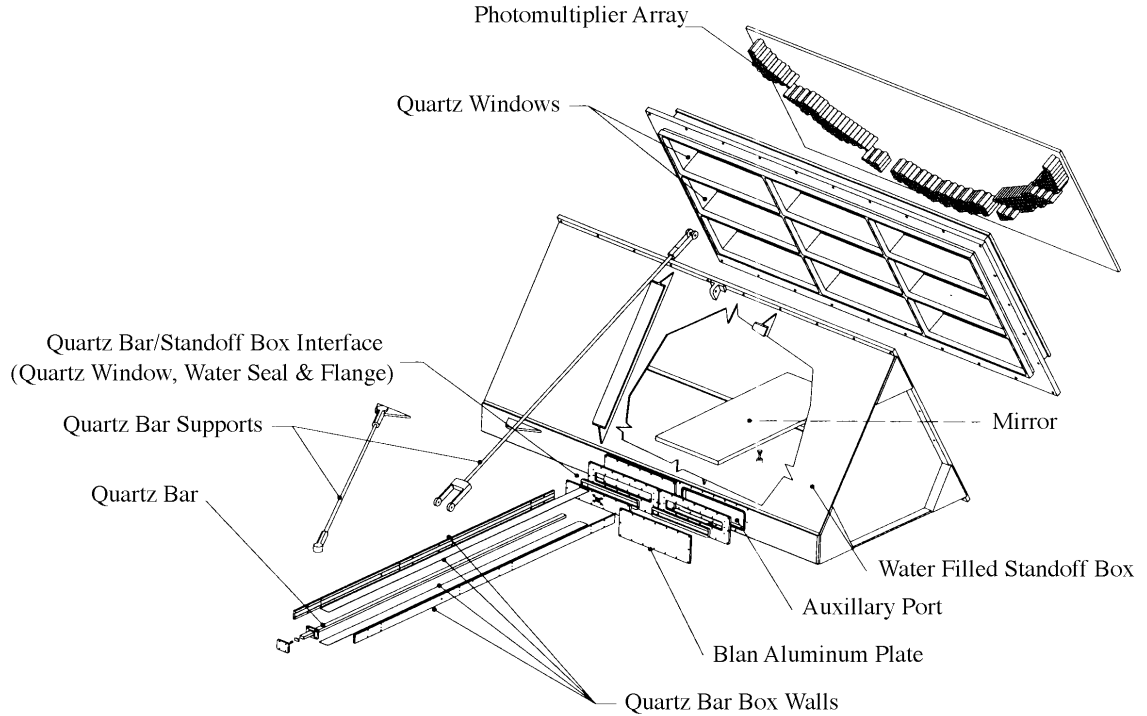,width=\textwidth}}
	\caption{\it \label{fig:proto2}
	Isometric drawing of the \dirc\ full scale prototype.}
  \end{center}
\end{figure}

\subsubsection{Setup}
\label{sec:p2setup}
The {\protoII} consisted of three
major mechanical assemblies (see Figure \ref{fig:proto2}),
installed on a carriage which could translate and
rotate {\protoII} in several dimensions:

\begin{itemize}
\item Two 120~cm long quartz bars (manufactured by
	Zygo Corp.~\cite{zygo} from Vitreosil~F/055 fused quartz
	material~\cite{qpc}), inside a protective box.
	The bars were placed in the box either side-by-side 
	or glued together as one 240~cm long bar.
	The bars were coupled to the {\SOB} through
	a small quartz entrance window.
\item A standoff box filled with pure water, with its mechanical support.
	This assembly included a small quartz window interface where {\cer}
	light from the bar entered the {\SOB}, and nine large quartz windows
	in a $3 \times 3$ array, which allowed the light to pass through
	to the {\PMT}s after image expansion.
	The bottom of the {\SOB} held a plane glass mirror with front-surface
	aluminum and dielectric coatings~\cite{OCI}.
	This mirror is used to reflect the lower \cer\ ring image onto the upper one,
	reducing the number of PMTs needed by 50\%.
\item An array of {\PMT}s (Hamamatsu model R268~\cite{hamamatsu}).
	These were optically and mechanically coupled to the {\SOB} exit
	windows through a thin UV-clear silicone rubber sheet.
	For economic reasons, instead of instrumenting the full image plane,
	only a limited number of {\PMT}s were installed that covered the
	{\cer} ring associated with a single particle incident angle.
\end{itemize}

The T9 beam line 
provided unseparated secondary particles in a narrow momentum range
tunable between 0.8 and $10 \gevc$.
It was operated with positive particles (mainly protons, pions, and
positrons) at low intensity, delivering around $10^4$ particles to the test
area per  spill.
For particle identification, the beam line was equipped with two gas
threshold {\cer} counters and a time-of-flight system. 
To measure the angle and position of particles incident on the quartz bar,
three multi-wire proportional chambers 
were installed on the beam line.
The trigger was generated as the coincidence of signals from several plastic
scintillation counters. 
A beam halo veto counter made of scintillation counters
was installed on the beam line close to the beam extraction.
A large scintillation counter was also fixed
on the {\SOB} to flag particles crossing the water tank.
Another set of counters was arranged along the quartz bar
outside the region populated by trigger particles.
These counters were used offline to tag events with
associated beam halo particles.

\begin{figure}[b]
  \begin{center}
    \mbox{\epsfig{figure=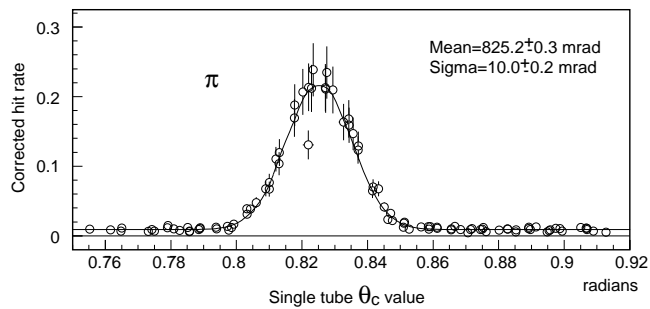,width=9.cm,clip=}}
  \caption
    {\it \label{fig:sing}
	{\cer} angle ({\thc}) distribution of the hit rate
	for pions, fit to a {\gau} signal
	plus flat background.  The fit result agrees well with
	the expected mean signal value of $\thc = 825.2 \mrad$
	for pions at $5.4 \gevc$ momentum.
    }
  \end{center}
\end{figure}

\subsubsection{Results}

The single photon {\thc} resolution of {\protoII} is
demonstrated in Figure \ref{fig:sing}.
This shows the distribution of corrected average hit rate versus
reconstructed {\thc} for a subset of {\PMT}s for pions
at $5.4 \gevc$, $\dip = - 20 \deg$, and $Z =  220 \cm$.
The data fit well to a {\gau} for the {\cer} signal
plus a flat distribution for the background. The {\thc}
resolution was defined as the {\gau} $\sigma$ parameter of this fit.
The single photon {\thc} resolution for pions was found to be
$10.0 \pm 0.2 \mrad$, in agreement with Monte Carlo simulations
and consistent with the expectation based on the value obtained for 
an air standoff region in {\protoI}.
Similar analysis of protons and electrons gave consistent results.
No significant variation in this resolution with either
$Z$ (photon transmission distance down the bar) or {\phc} (photon position
within the {\cer} ring) was observed, within the calculated errors of the data.

The particle identification power of {\protoII} was studied by measuring
the average \cer\ angle per track for pions and protons at $5.4 \gevc$.
At that momentum, the {\thc} separation between pion and
proton {\cer} rings has the same value
as between pions and kaons at $2.8 \gevc$.
The momentum $2.8 \gevc$ corresponds roughly to that of decay products from a
two-body B$^0$ decay in {\babar} hitting the {\dirc} bar
at the same $20^{\circ}$ value of {\dip}.
In this decay mode,
pion/kaon separation is essential for the physics goal of the
{\babar} experiment.  
Thus, even though the {\babar} {\dirc} detector differs in detail from 
the {\protoII} detector, 
the pion/proton separation at $5.4 \gevc$ tests the performance of 
{\protoII} in a situation relevant for the {\babar} \dirc .

\begin{figure}[b]
  \begin{center}
    \mbox{\epsfig{figure=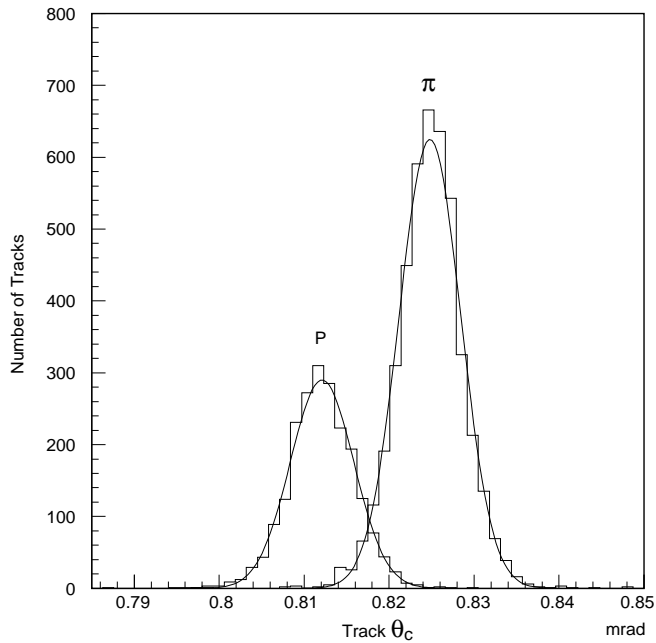,width=9.cm}}
  \caption
    {\it \label{fig:pipr}
	{\cer} angle per track for pions and protons
	at $5.4 \gevc$, with  $Z = 220 \cm$.
    }
  \end{center}
\end{figure}

The particle separation power was defined as the
difference in the {\thc} peak position values,
divided by the {\rms} of the pion
track {\thc} distribution.  This is shown in Figure \ref{fig:pipr}.
The pion track {\thc} {\rms} was measured
for data at $Z = 220\cm$, and found to be $3.6 \mrad$.
The peak separation was measured in the same data to be $13 \mrad$, consistent
with expectations.
These together give a pion/proton PID power for {\protoII} of
3.6 standard deviations.

The track \cer\ angle resolution was observed to
scale approximately as the square root of the number of photons, with
a small residual term of about $1.1 \mrad$ coming from correlated
factors such as track multiple scattering in the bar, and residual
alignment uncertainties.

\begin{figure}
  \begin{center}
    \mbox{\epsfig{figure=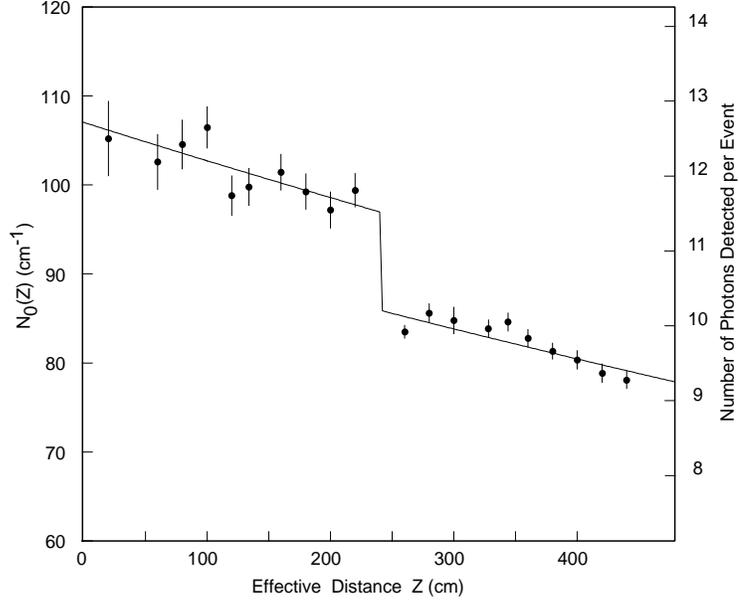,width=10.cm}}
  \caption
    {\it \label{fig:nvsz}
	$N_0$ as a function of the effective bar transmission distance $Z$
	for the coarse $Z$ scan
	at $\dip=\pm 20\deg$
	and $5.4 \gevc$ momentum.  Pion and proton samples are averaged in the
	data points shown.
	The discontinuity at $Z=240\cm$ accounts for the end mirror reflectivity.
	A possible loss at the glue joint (for $Z=120\cm$ and $Z=360\cm$) was
	not included in this fit.  The scale at the right indicates the 
	average number of signal photons extracted from a ring fit
	in the range $|\phc| < 1.0 \rad$.
    }
  \end{center}
\end{figure}

The {\photoelectron} yield of {\protoII} was measured with a coarse
$Z$ scan of the 240~cm long bar with $Z$ steps of $20\cm$ over
the full $Z$ range (20~cm to 460~cm) at constant momentum
(5.4 {\gevc}) and particle incident angle ($\pm 20\deg$).
This analysis also allows the determination of the
attenuation rate for {\cer} photons as they traveled down the bar.
The {\cer} quality factor $N_0$
is defined by the relation
$\epsilon_{geom} \times N_0 \times \sin^2{\theta_C} =$
the number of {\photoelectron}s detected per centimeter of radiator
over the acceptance of {\protoII}.
This definition of $N_0$ differs slightly from the one conventionally used
with threshold {\cer} detectors~\cite{PDG} in that the geometric acceptance
$\epsilon_{geom}$, which accounts for the truncation of the
{\cer} cone and for the partial coverage of the image plane
by {\PMT}s, has been explicitly extracted.
In this definition, $N_0$ is proportional to the product of the
photodetection efficiency and the collection efficiency $\epsilon_{coll}$,
which accounts for
the light losses during photon transmission and reflection.
For the {\dirc , $N_0$ is a function of $Z$.

Figure~\ref{fig:nvsz} shows the measurements of $N_0$($Z$) versus $Z$.
The data are fit to an exponential with a discontinuity
at $Z=240 \cm$, to account for reflection loss at the bar-end mirror.
The fit gives a $\chi^2$ of 24 for 18 degrees of freedom, and the
fitted values for the three parameters are
\begin{center}
\begin{tabular}{rcrcl}
$N_0(0)$  &=& $106.9$ &$\pm$& $1.3$ cm$^{-1}$ \\
attenuation per meter  &=&  $4.1$  &$\pm$& $0.7$\%\\
loss at the mirror     &=& $11.4$  &$\pm$& $1.5$\%  \\
\end{tabular}
\end{center}

The errors quoted above are those returned from the
fit, and so include the full statistical and systematic error on
the photon rates, but not any systematic error due to the
fit model itself.
In this fit the potential loss of light at the glue joint at $Z=120 \cm$
(and $Z=360 \cm$) is ignored.
Fits with this loss as a free parameter were also performed.
The glue joint light loss could not be measured precisely and was found to be
consistent with zero within two standard deviations.
Lab measurements and other tests also indicate a negligible light loss
at the glue joint.
Neglecting the glue joint loss leads to a conservative (larger)
estimate of the attenuation, and has no effect on the estimate of $N_0 $(0).
To facilitate comparing this $N_0(0)$ measurement with other
$N_0$ measurements, the effects peculiar to the {\protoII} test
setup implicit in $\epsilon_{coll}$ must be removed.
There is no way to measure $\epsilon_{coll}$ directly from the data.
However, a realistic estimate of $\epsilon_{coll}$ at $Z=0$ can be
calculated by combining expected light-loss
effects from the different parts of {\protoII},
resulting in the estimate $\epsilon_{coll}
\simeq 73\%$ at $Z=0$, which leads to:
\[
N_0(0)/\epsilon_{coll} = 146  \pm 1.8 \pm 9 \,\mbox{cm}^{-1}
\label{n0_result}
\]
The first error comes from the
$N_0$ fit error, the second from an estimated 30\% uncertainty
in the light loss estimates.
This value is in agreement with the value of $137 \,\mbox{cm}^{-1}$ that
was used for performance estimates of the {\babar} {\dirc} in the {\babar}
TDR~\cite{tdr} and the {\protoI} result of $150 \,\mbox{cm}^{-1}$.

This measurement of the attenuation value agrees well with a
full Monte Carlo simulation, which
includes the spectral response of the {\PMT}s, the spectral transmission of
bulk fused quartz, and benchtop measurements~\cite{dn18,dn_trans}.
It is considerably lower than the value of 10\% observed in {\protoI}.
The difference can be understood as the result of surface contamination.
Careful cleaning of the bar surfaces was shown to reduce the attenuation
from 10\% to about 4\%.

Other results included:
\begin{itemize}
\item 	Scans across the gap region between bars in the side-by-side operation
	of two 120~cm long bars showed that there is no measurable cross talk
	between bars mounted in close proximity side-by-side.
\item 	The effective index of refraction of the quartz
	radiator was measured to be $n_q=1.474$ from a direct fit to the momentum
	dependent {\cer} angle.
\item	One of the large quartz windows that coupled the SOB to the PMT array
	was replaced with a jig which held the {\PMT}s directly in the water.
	The operation of the photomultiplier tubes in water was completely
	satisfactory.
\item   The {\babar} {\dirc} proposal to equip each {\PMT}
	with light concentrators was tested by equipping one of the PMT
	bundles temporarily with a set of reflective cones.
	Each cone was $8\mm$ tall with an outer diameter of $31\mm$ equal
	to the {\PMT} spacing, and an inner diameter of $25\mm$ equal to
	the photocathode diameter.
	To facilitate their mounting, ten cones were manufactured together
	in a single plaque of molded plastic.
	The cone plaques were polished and coated with a $600\nm$ Al layer
	followed by $200\nm$ of SiO$_2$.
	The light collection improvement using the concentrators
	measured in the {\protoII} test was consistent with lab
	measurements.
\item 	Finally, the relationship between timing and photon
	position at the detector was shown to be very useful for background
	rejection.
	The single photon timing resolution was measured to be $2.4\,\ns$
	and was dominated, as expected, by the transit time spread of the PMTs
	used in the test.
	The {\babar} {\dirc} will use {\PMT}s with
	a $1.8 \ns$ transit time spread (ETL model 9125~\cite{emi_tubes}).
\end{itemize}

In conclusion, the operation of {\protoII} was stable and robust over
a periode of 12 months in the T9 beam.
The tests were very successful and no significant, unanticipated
variance in performance as a function of the position or angle of the
track in the {\dirc} bar was observed.

\section{The Fused Silica Radiators}
\label{sec:quartz}

The optical properties and radiation hardness of several quartz radiator 
material candidates were studied in a program of research and development 
that aimed at determining whether bars that meet the stringent requirements 
of the \dirc\ can be obtained from commercial sources.

The radiation hardness was tested by comparing the transmission of 
several samples of ``natural'' fused quartz (made from crushing and melting 
natural crystals or sand) and synthetic fused silica (made by flame 
hydrolysis of Silicon-bearing compounds, such as ${\rm SiCl_4}$) 
before and after exposure to doses of up to 500 krad 
from a ${\rm Co^{60}}$ source (1.17/1.33 MeV photons).
All natural quartz materials, inculding the Vitreosil~F/055 used in the 
prototypes, were found to show unacceptable transmission losses after only
a few krad exposure.
The synthetic types, on the other hand, showed little or no transmission loss,
even after exposure to several hundred krad~\cite{dn18,dn39}.
A detailed comparison of different types of synthetic materials 
revealed that a periodicity in the refractive index
of the quartz bulk material, a by-product of the proprietary production
process, causes interference effects in some synthetic quartz 
materials~\cite{dn_lobe}.
This interference would severly decrease the angular resolution of the \dirc .
A candidate material that does not show the periodicity was identified
(Spectrosil~\cite{qpc}) and will be used in the \dirc .

In order to preserve the photon angles during surface reflections,
the faces and sides have to be nominally parallel while the orthogonal
surfaces are kept nominally perpendicular.
Typically, the bar's surfaces have to be flat and parallel to
about 25 $\mu$m, while the orthogonal surfaces are perpendicular to a
tolerance of 0.3 mrad.
The most difficult requirements are associated with maintaining the photon
transmission during reflections at the surfaces of the bar (a \cer\ photon
may be internally reflected a few hundred times before exiting the
bar).
This leads to rather severe requirements on edge sharpness and surface finish.
After polishing, the \dirc\ radiator bars have an average edge radius less than
5 $\mu$m, and a nominal surface polish of better than 0.5 nm (RMS).
Benchtop measurements, using a polarized HeCd laser beam at 442~nm, 
demonstrated that the coefficient of total internal reflection of bars polished
to this high quality finish exceeds 0.9995~\cite{dn_trans}.
The bars will be manufactured by Boeing~\cite{boeing}.

\section{The {\dirc} for {\babar}}
\label{sec:babar_dirc}

{\protoI} and {\protoII} have shown that the {\dirc} is well-matched to the
asymmetric {\BF} and motivate the choice of the \dirc\ as the primary PID
system for \babar .
Simulations based on the prototype performance predict an excellent $\pi/K$
separation of the \babar\ \dirc .
This is demonstrated in Figure~\ref{fig:nsig} where the predicted PID
performance is shown as a function of the particle momentum and the polar angle
$\cos\theta$.
The separation is nearly four standard deviations or better over the entire
acceptance region.

\begin{figure}[b]
  \begin{center}
    \mbox{\epsfig{figure=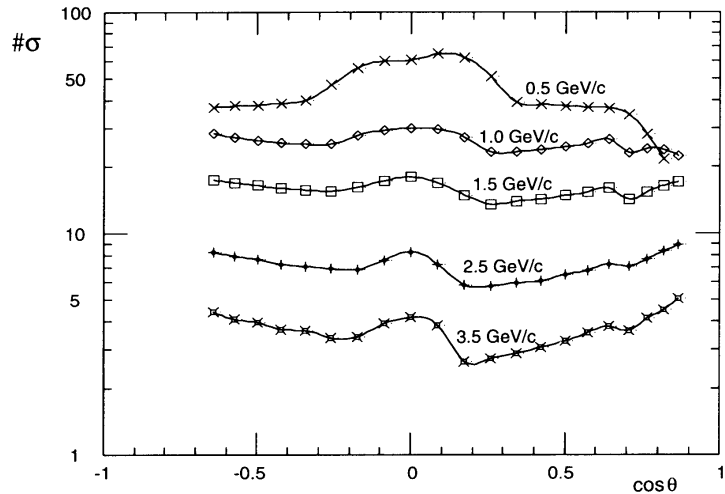,width=11.cm}}
  \caption
    {\it \label{fig:nsig}
	Predicted $\pi$/K separation performance of the \dirc ,
	quoted in terms of the number of standard deviations,
	\vs\ \ctheta, for different momenta.
    }
  \end{center}
\end{figure}

The \dirc\ for \babar\ will be a barrel detector, located in radius between the
drift chamber and the CsI calorimeter.
The main components are shown schematically in Figure~\ref{fig:comp}.

\begin{figure}[t]
  \begin{center}
    \mbox{\epsfig{figure=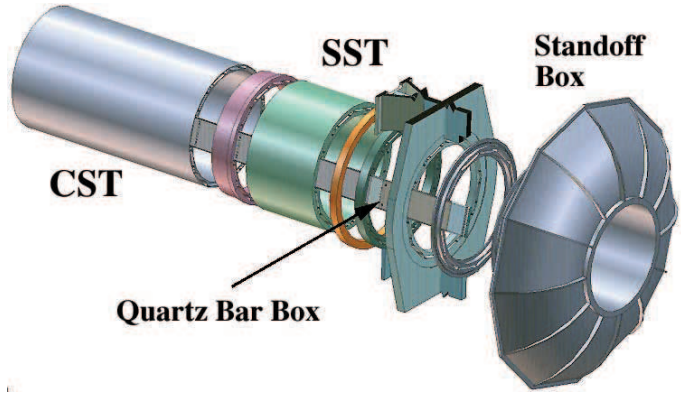,width=9.cm}}
  \caption
    {\it \label{fig:comp}
	Schematic of the \dirc\ detector assembly.
    }
  \end{center}
\end{figure}

\begin{figure}[b]
  \begin{center}
    \hspace*{-2mm}
    \mbox{\epsfig{figure=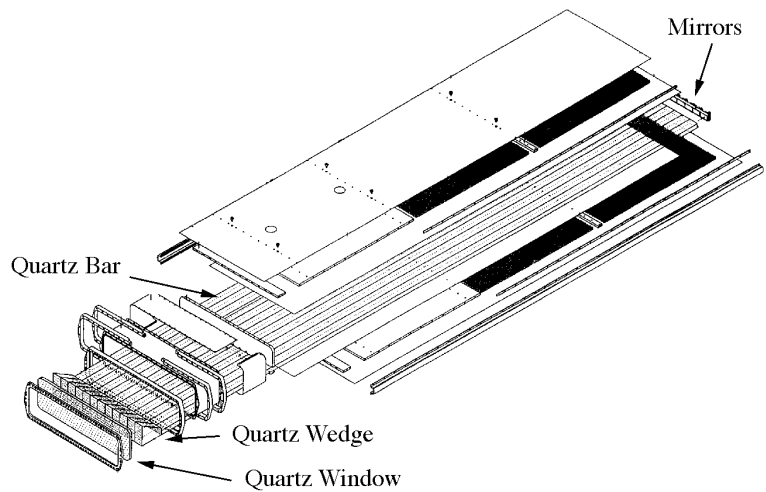,width=11cm}}
  \caption
    {\it \label{fig:bab}
	Schematic of the \dirc\ bar box assembly.
    }
  \end{center}
\end{figure}

The mechanical support of the \dirc\ is cantilevered from the iron endcap
region, supported by a thick steel tube (SST) which
also helps to minimize the magnetic flux gap caused by the \dirc\ extending
through the instrumented flux return.
The quartz radiator bars will be supported in the active region by
a thin extension of this tube.
This central support tube (CST) is designed similar to an aircraft wing,
made of thin aluminum inner and outer shells covering an aluminum frame.
The frame will be made
of thin-wall bulkhead rings spaced every $60\cm$ along the tube axis.
The space between the bulkheads will be filled by
construction foam.  There will be no {\dirc} mechanical supports in the
forward end of {\babar}, minimizing its impact on the other detector
systems located there.
The standoff box will be made of stainless steel and hold about
6\,000 liters of pure water.
The {\babar} {\dirc} will use almost 11\,000 $2.82\cm$ diameter
{\PMT}s (ETL model 9125~\cite{emi_tubes}).
These tubes have high gain and good quantum efficiency (around 25\%)
in the {\cer} wavelengths transmitted by both quartz and water, and
are available at modest price.
They will be arranged in a nearly close-packed hexagonal pattern
on the detection surface.
Hexagonal light concentrators on the front of the {\PMT}s will
result in an effective
active surface area fraction of approximately 90\%.
The {\PMT}s will lie on a surface that is approximately toroidal.
The distance
traversed in the water by the photons emerging from the bar end
will be $1.17\m$.
This distance, together with the size of the bars and {\PMT}s, gives a
geometric contribution to the single photon {\cer} angle resolution of
$ 7\mrad$.
This geometric contribution is approximately equal to the resolution
contribution coming from the production and transmission dispersions.

The total thickness of the \dirc , including support structures, will be about
8~cm or 19\% of a radiation length for a particle at normal incidence.
The {\dirc} bars will be arranged as a 12-sided polygonal barrel.
Each side of the polygon will consist of 12 bars placed
very close together ($75 \mum$ gap) side by side, in 12 quartz bar boxes,
shown in Figure~\ref{fig:bab}, for a total of 144 bars.
The radiator bars will cover 94\% of the azimuthal angle and 87\%
of the center-of-mass polar angle.
The bars will be made from synthetic fused silica and will 
have transverse dimensions of $1.7\cm$ thick by 3.5\,cm wide,
and are about $4.90\m$ long.
The length is achieved by gluing end-to-end four $1.225\m$ bars, that size
being the longest high quality quartz bar currently available from industry.

The mirror in the {\SOB} described in the section on the {\protoII} setup,
has been replaced in the {\babar} {\dirc} design by
a quartz ``wedge'' which is glued to the readout end of each bar.
The wedge is a $9\cm$ long block of synthetic fused silica with the same
width as the bars ($3.5\cm$), and a trapezoidal profile ($2.8 \cm$ high at
the bar end and $8\cm$ high at the quartz window, which provides the
interface to the water).
Total internal reflection on all sides of the quartz wedge provides nearly
lossless reflection, thereby increasing the number of detectable photons
relative to the {\protoII} mirror design.
The wedge design also slightly improves the angular resolution,
allows a stronger and more robust water seal, and eliminates
the need for operating the fragile mirrors in the standoff box water
and their post-installation alignment.

\section{Conclusions}

The \dirc\ is a new type of ring imaging \cer\ detector that
is well-matched to the requirements for a particle identification device
in the BaBar detector at the {\PEPII} {\BF}.
It is thin, fast, and tolerant of background.
The prototype program has demonstrated that the principles of operation
are well-understood, and that the final system is expected to yield
excellent $\pi/K$ separation of nearly four standard deviations
or better over the full kinematic region for all of the
products from B decays.
The construction of the \dirc\ is well underway.
The detector will be installed in \babar\ by September 1998.
This will be followed by a cosmic ray checkout and data taking is
scheduled to begin in April 1999.

\section{Acknowledgments}

This research is supported by the Department of Energy under contracts
DE-AC03-76SF00515 (SLAC), DE-AC03-76SF00098 (LBNL), DE-AM03-76SF0010
(UCSB), and DE-FG03-93ER40788 (CSU); the National Science Foundation
grants PHY-95-10439 (Rutgers) and PHY-95-11999 (Cincinnati).  
The speaker would like to thank the Alex\-ander-\-von-\-Hum\-boldt
Stiftung for their financial support.


\begin{thebibliography}{9}

\bibitem{pep2}
``An Asymmetric B Factory Based on PEP: Conceptual Design Report,''
{\em LBL-PUB-5303/SLAC-REP-372 (1991)}.

\bibitem{tdr}
The \babar\ Collaboration,
``\babar\ Technical Design Report,''
{\em SLAC-REP-950457, 1995}.

\bibitem{dirc_concept}
B.N. Ratcliff, 
``The B Factory Detector for \PEPII : a Status Report,''
SLAC-PUB-5946 (1992) and Dallas HEP (1992) 1889; \\
B.N. Ratcliff, 
``The \dirc\ Counter: a New Type of Particle Identification Device for 
B Factories,''
SLAC-PUB-6047 (1993); \\
P. Coyle et.al.,
``The \dirc\ Counter: A New Type of Particle Identification
Device for B Factories,''
{\em Nucl. Instr. Methods~ A 343 1994} pp. 292. 

\bibitem{hide_ieee}
D. Aston et.al.,
{ ``Test of a Conceptual Prototype of the Total Internal
Reflection \cer\ Imaging Detector (\dirc ) with Cosmic Muons,''}\\
{{\sl IEEE Trans. Nucl. Sci.~}} 42 (1995) 534.

\bibitem{nim_paper}
H. Staengle et.al.,
``Test of a Large Scale Prototype of the {\sc Dirc}, a
\cer\ Imaging Detector based on Total Internal Reflection for
{\sc BaBar}  at {\sc Pep-II},''\\
{\sl Nucl. Instr. Methods~} A 397 (1997) 261.

\bibitem{zygo}
Zygo Corporation, Laurel Brook Road,
Middlefield, Connecticut 06455.

\bibitem{qpc}
Quartz Products Co., 1600 W. Lee St., Louisville, Kentucky 40201.

\bibitem{epotek}
Epoxy Technology, Inc.,
14 Fortune Dr., Billerica, Massachusettes 01821.

\bibitem{pmtp1}
Burle Industries, Inc.,
1000 New Holland Ave.,
Lancaster, Pennsylvania 17601.

\bibitem{emi_tubes}
Electron Tubes Limited (formerly: Thorn EMI Electron Tubes),
Bury Street, Ruislip,
Middlesex HA47TA, U.K.

\bibitem{OCI}Optical Coating Laboratory Inc. (OCLI)
2789 Northpoint Pkwy, Santa Rosa, California 95407.

\bibitem{hamamatsu}
Hamamatsu Co., 
250 Wood Ave, 
Middlesex, New Jersey 08846.

\bibitem{PDG}
Particle Data Group,
``Review of Particle Physics,''
{\sl Physical Review {D 54}, Part 1 (1996)}.

\bibitem{dn18}
H. Krueger,  M. Schneider, R.  Reif, J. Va'vra, 
``Initial Measurements of Quartz Transmission, Internal Reflection
Coefficient and the Radiation Damage,''
{\em Internal \dirc\ Note \#18, January 1996}.

\bibitem{dn_trans} 
H. Krueger, R. Reif, X. Sarazin, J. Schwiening, J. Va'vra,
``Measuring the Optical Quality of Quartz Bars and the Coupling of RTV
to the Window,''
{\em Internal \dirc\ Note \#40, May 1996};\\
H. Krueger, R. Reif, X. Sarazin, J. Schwiening and J. Va'vra,
``The Optical Scanning System for the Quartz Bar Quality Control,''
{\em Internal \dirc\ Note \#54, October 1996}.

\bibitem{dn39}
X. Sarazin, M. Schneider, J. Schwiening, R. Reif and
J. Va'vra,
``Radiation Damage of Synthetic Quartz and Optical Glues,''
{\em Internal \dirc\ Note \#39, May 1996}.

\bibitem{dn_lobe} 
M. Convery, B. Ratcliff, J. Schwiening and J. Va'vra,
``Measurements of Periodic Structure in Synthetic Quartz,''
{\em Internal \dirc\ Note \#87, September 1997}.

\bibitem{boeing}
Boeing,
2511 C Broadbent Parkway NE,
Albuquerque, New Mexico 87107.

\end{thebibliography}
\end{document}